\documentclass[10pt,a4paper,onecolumn]{article}
\usepackage{marginnote}
\usepackage{graphicx}
\usepackage{xcolor}
\usepackage{authblk,etoolbox}
\usepackage{titlesec}
\usepackage{calc}
\usepackage{tikz}
\usepackage{hyperref}
\hypersetup{colorlinks,breaklinks,
            urlcolor=[rgb]{0.0, 0.5, 1.0},
            linkcolor=[rgb]{0.0, 0.5, 1.0}}
\usepackage{caption}
\usepackage{tcolorbox}
\usepackage{amssymb,amsmath}
\usepackage{ifxetex,ifluatex}
\usepackage{seqsplit}
\usepackage{xstring}

\usepackage{float}
\let\origfigure\figure
\let\endorigfigure\endfigure
\renewenvironment{figure}[1][2] {
    \expandafter\origfigure\expandafter[H]
} {
    \endorigfigure
}

\usepackage{fixltx2e} % provides \textsubscript
\usepackage[
  backend=biber,
]{biblatex}
\bibliography{paper.bib}

\let\textttOrig=\texttt
\def\texttt#1{\expandafter\textttOrig{\seqsplit{#1}}}
\renewcommand{\seqinsert}{\ifmmode
  \allowbreak
  \else\penalty6000\hspace{0pt plus 0.02em}\fi}

\makeatletter
\let\href@Orig=\href
\def\href@Urllike#1#2{\href@Orig{#1}{\begingroup
    \def\Url@String{#2}\Url@FormatString
    \endgroup}}
\def\href@Notdoi#1#2{\def\tempa{#1}\def\tempb{#2}%
  \ifx\tempa\tempb\relax\href@Urllike{#1}{#2}\else
  \href@Orig{#1}{#2}\fi}
\def\href#1#2{%
  \IfBeginWith{#1}{https://doi.org}%
  {\href@Urllike{#1}{#2}}{\href@Notdoi{#1}{#2}}}
\makeatother

\usepackage[top=3.5cm, bottom=3cm, right=1.5cm, left=1.0cm,
            headheight=2.2cm, reversemp, includemp, marginparwidth=4.5cm]{geometry}

\titleformat{\section}
  {\normalfont\sffamily\Large\bfseries}
  {}{0pt}{}
\titleformat{\subsection}
  {\normalfont\sffamily\large\bfseries}
  {}{0pt}{}
\titleformat{\subsubsection}
  {\normalfont\sffamily\bfseries}
  {}{0pt}{}
\titleformat*{\paragraph}
  {\sffamily\normalsize}

\usepackage{fancyhdr}
\makeatletter
\let\ps@plain\ps@fancy
\fancyheadoffset[L]{4.5cm}
\fancyfootoffset[L]{4.5cm}

\definecolor{linky}{rgb}{0.0, 0.5, 1.0}

\newtcolorbox{repobox}
   {colback=red, colframe=red!75!black,
     boxrule=0.5pt, arc=2pt, left=6pt, right=6pt, top=3pt, bottom=3pt}

\newcommand{\ExternalLink}{%
   \tikz[x=1.2ex, y=1.2ex, baseline=-0.05ex]{%
       \begin{scope}[x=1ex, y=1ex]
           \clip (-0.1,-0.1)
               --++ (-0, 1.2)
               --++ (0.6, 0)
               --++ (0, -0.6)
               --++ (0.6, 0)
               --++ (0, -1);
           \path[draw,
               line width = 0.5,
               rounded corners=0.5]
               (0,0) rectangle (1,1);
       \end{scope}
       \path[draw, line width = 0.5] (0.5, 0.5)
           -- (1, 1);
       \path[draw, line width = 0.5] (0.6, 1)
           -- (1, 1) -- (1, 0.6);
       }
   }

\patchcmd{\@maketitle}{center}{flushleft}{}{}
\patchcmd{\@maketitle}{center}{flushleft}{}{}
\patchcmd{\@maketitle}{\LARGE}{\LARGE\sffamily}{}{}
\def\maketitle{{%
  
  \AB@maketitle}}
\makeatletter
\renewcommand\AB@affilsepx{ \protect\Affilfont}
\renewcommand\AB@affilnote[1]{{\bfseries #1}\hspace{3pt}}
\renewcommand{\affil}[2][]%
   {\newaffiltrue\let\AB@blk@and\AB@pand
      \if\relax#1\relax\def\AB@note{\AB@thenote}\else\def\AB@note{#1}%
        \setcounter{Maxaffil}{0}\fi
        \begingroup
        \let\href=\href@Orig
        \let\texttt=\textttOrig
        \let\protect\@unexpandable@protect
        \def\thanks{\protect\thanks}\def\footnote{\protect\footnote}%
        \@temptokena=\expandafter{\AB@authors}%
        {\def\\{\protect\\\protect\Affilfont}\xdef\AB@temp{#2}}%
         \xdef\AB@authors{\the\@temptokena\AB@las\AB@au@str
         \protect\\[\affilsep]\protect\Affilfont\AB@temp}%
         \gdef\AB@las{}\gdef\AB@au@str{}%
        {\def\\{, \ignorespaces}\xdef\AB@temp{#2}}%
        \@temptokena=\expandafter{\AB@affillist}%
        \xdef\AB@affillist{\the\@temptokena \AB@affilsep
          \AB@affilnote{\AB@note}\protect\Affilfont\AB@temp}%
      \endgroup
       \let\AB@affilsep\AB@affilsepx
}
\makeatother

\renewcommand\Affilfont{\sffamily\small\mdseries}
\ifnum 0\ifxetex 1\fi\ifluatex 1\fi=0 % if pdftex
  \usepackage[T1]{fontenc}
  \usepackage[utf8]{inputenc}

\else % if luatex or xelatex
  \ifxetex
    \usepackage{mathspec}
  \else
    \usepackage{fontspec}
  \fi
  \defaultfontfeatures{Ligatures=TeX,Scale=MatchLowercase}

\fi
\IfFileExists{upquote.sty}{\usepackage{upquote}}{}
\IfFileExists{microtype.sty}{%
\usepackage{microtype}
\UseMicrotypeSet[protrusion]{basicmath} % disable protrusion for tt fonts
}{}

\usepackage{hyperref}
\hypersetup{unicode=true,
            pdftitle={GGLasso - a Python package for General Graphical Lasso computation},
            pdfborder={0 0 0},
            breaklinks=true}
\usepackage{color}
\usepackage{fancyvrb}

\DefineVerbatimEnvironment{Highlighting}{Verbatim}{commandchars=\\\{\}}
\newenvironment{Shaded}{}{}

\newcommand{\CommentTok}[1]{\textcolor[rgb]{0.38,0.63,0.69}{\textit{#1}}}

\newcommand{\FloatTok}[1]{\textcolor[rgb]{0.25,0.63,0.44}{#1}}

\newcommand{\ImportTok}[1]{#1}

\newcommand{\NormalTok}[1]{#1}
\newcommand{\OperatorTok}[1]{\textcolor[rgb]{0.40,0.40,0.40}{#1}}

\newcommand{\StringTok}[1]{\textcolor[rgb]{0.25,0.44,0.63}{#1}}
\newcommand{\VariableTok}[1]{\textcolor[rgb]{0.10,0.09,0.49}{#1}}

\usepackage{longtable,booktabs}

\let\addcontentslineOrig=\addcontentsline
\def\addcontentsline#1#2#3{\bgroup
  \let\texttt=\textttOrig\addcontentslineOrig{#1}{#2}{#3}\egroup}
\let\markbothOrig\markboth
\def\markboth#1#2{\bgroup
  \let\texttt=\textttOrig\markbothOrig{#1}{#2}\egroup}
\let\markrightOrig\markright
\def\markright#1{\bgroup
  \let\texttt=\textttOrig\markrightOrig{#1}\egroup}

\usepackage{graphicx,grffile}
\makeatletter
\def\maxwidth{\ifdim\Gin@nat@width>\linewidth\linewidth\else\Gin@nat@width\fi}
\def\maxheight{\ifdim\Gin@nat@height>\textheight\textheight\else\Gin@nat@height\fi}
\makeatother
\setkeys{Gin}{width=\maxwidth,height=\maxheight,keepaspectratio}
\IfFileExists{parskip.sty}{%
\usepackage{parskip}
}{% else
\setlength{\parindent}{0pt}
\setlength{\parskip}{6pt plus 2pt minus 1pt}
}
\providecommand{\tightlist}{%
  \setlength{\itemsep}{0pt}\setlength{\parskip}{0pt}}
\ifx\paragraph\undefined\else
\let\oldparagraph\paragraph
\renewcommand{\paragraph}[1]{\oldparagraph{#1}\mbox{}}
\fi
\ifx\subparagraph\undefined\else
\let\oldsubparagraph\subparagraph
\renewcommand{\subparagraph}[1]{\oldsubparagraph{#1}\mbox{}}
\fi

\title{GGLasso - a Python package for General Graphical Lasso computation}

        \author[1]{Fabian Schaipp}
          \author[2,3]{Oleg Vlasovets}
          \author[2,3,4]{Christian L. Müller}
    
      \affil[1]{Technische Universität München}
      \affil[2]{Institute of Computational Biology, Helmholtz Zentrum München}
      \affil[3]{Department of Statistics, Ludwig-Maximilians-Universität München}
      \affil[4]{Center for Computational Mathematics, Flatiron Institute, New York}
  \date{\vspace{-5ex}}

\begin{document}
\maketitle

\marginpar{
  \sffamily\small

  {\bfseries DOI:} \href{https://doi.org/}{\color{linky}{}}

  \vspace{2mm}

  {\bfseries Software}
  \begin{itemize}
    \setlength\itemsep{0em}
    \item \href{}{\color{linky}{Review}} \ExternalLink
    \item \href{https://github.com/fabian-sp/GGLasso}{\color{linky}{Repository}} \ExternalLink
    \item \href{}{\color{linky}{Archive}} \ExternalLink
  \end{itemize}

  \vspace{2mm}

  {\bfseries Submitted:} 18 October 2021\\
  {\bfseries Published:} 

  \vspace{2mm}
  {\bfseries License}\\
  Authors of papers retain copyright and release the work under a Creative Commons Attribution 4.0 International License (\href{https://creativecommons.org/licenses/by/4.0/}{\color{linky}{CC BY 4.0}}).
}

\hypertarget{summary}{%
\section{Summary}\label{summary}}

We introduce \texttt{GGLasso}, a Python package for solving General
Graphical Lasso problems. The Graphical Lasso scheme, introduced by
(Friedman, Hastie, and Tibshirani 2007) (see also (Yuan and Lin 2007;
Banerjee, El Ghaoui, and D'Aspremont 2008)), estimates a sparse inverse
covariance matrix \(\Theta\) from multivariate Gaussian data
\(\mathcal{X} \sim \mathcal{N}(\mu, \Sigma) \in \mathbb{R}^p\).
Originally proposed by (Dempster 1972) under the name Covariance
Selection, this estimation framework has been extended to include latent
variables in (Chandrasekaran, Parrilo, and Willsky 2012). Recent
extensions also include the joint estimation of multiple inverse
covariance matrices, see, e.g., in (Danaher, Wang, and Witten 2013;
Tomasi et al. 2018). The \texttt{GGLasso} package contains methods for
solving the general problem formulation:

\begin{align}
\label{eq:problem}
\min_{\Theta, L \in \mathbb{S}_{++}^K }\quad \sum_{k=1}^{K} \left(-\log\det(\Theta^{(k)} - L^{(k)}) + \langle S^{(k)},  \Theta^{(k)} - L^{(k)} \rangle \right)+ \mathcal{P}(\Theta) +\sum_{k=1}^{K} \mu_{1,k} \|L^{(k)}\|_{\star}.
\end{align}

Here, we denote with \(\mathbb{S}_{++}^K\) the \(K\)-product of the
space of symmetric, positive definite matrices. Moreover, we write
\(\Theta = (\Theta^{(1)},\dots,\Theta^{(K)})\) for the sparse component
of the inverse covariances and \(L = (L^{(1)},\dots,L^{(K)})\) for the
low rank components, formed by potential latent variables. Here,
\(\mathcal{P}\) is a regularization function that induces a desired
sparsity structure. The above problem formulation subsumes important
special cases, including the single (latent variable) Graphical Lasso,
the Group, and the Fused Graphical Lasso.

\hypertarget{statement-of-need}{%
\section{Statement of need}\label{statement-of-need}}

Currently, there is no Python package available for solving general
Graphical Lasso instances. The standard single Graphical Lasso problem
(SGL) can be solved in \texttt{scikit-learn} (Pedregosa et al. 2011).
The \texttt{skggm} package provides several algorithmic and model
selection extensions for the single Graphical Lasso problem (Laska and
Narayan 2017). The package \texttt{regain} (Tomasi et al. 2018)
comprises solvers for single and Fused Graphical Lasso problems, with
and without latent variables. With \texttt{GGLasso}, we make the
following contributions:

\begin{itemize}
\tightlist
\item
  Proposing a uniform framework for solving Graphical Lasso problems.
\item
  Providing solvers for Group Graphical Lasso problems (with and without
  latent variables).
\item
  Providing a solver for -- what we call -- \emph{nonconforming GGL}
  problems where not all variables need to be present in every instance.
  We detail a use case of this novel extension on synthetic data.
\item
  Implementing a block-wise ADMM solver for SGL problems following
  (Witten, Friedman, and Simon 2011) as well as proximal point solvers
  for FGL and GGL problems (Zhang et al. 2019, 2020).
\end{itemize}

In the table below we give an overview of existing functionalities and
the \texttt{GGLasso} package.

\begin{longtable}[]{@{}lllll@{}}
\toprule
& scikit-learn & regain & GGLasso & comment\tabularnewline
\midrule
\endhead
SGL & \textbf{yes} & \textbf{yes} & \textbf{yes} & new: block-wise
solver\tabularnewline
SGL + latent & \textbf{no} & \textbf{yes} & \textbf{yes}
&\tabularnewline
GGL & \textbf{no} & \textbf{no} & \textbf{yes} &\tabularnewline
GGL + latent & \textbf{no} & \textbf{no} & \textbf{yes} &\tabularnewline
FGL & \textbf{no} & \textbf{yes} & \textbf{yes} & new: proximal point
solver\tabularnewline
FGL + latent & \textbf{no} & \textbf{yes} & \textbf{yes}
&\tabularnewline
GGL nonconforming (+latent) & \textbf{no} & \textbf{no} & \textbf{yes}
&\tabularnewline
\bottomrule
\end{longtable}

\hypertarget{functionalities}{%
\section{Functionalities}\label{functionalities}}

\hypertarget{installation-and-problem-instantiation}{%
\subsection{Installation and problem
instantiation}\label{installation-and-problem-instantiation}}

\texttt{GGLasso} can be installed via \texttt{pip}.

\begin{verbatim}
pip install gglasso
\end{verbatim}

The central object of \texttt{GGLasso} is the class
\texttt{glasso\_problem} which streamlines the solving or model
selection procedure for SGL, GGL, and FGL problems with or without
latent variables.

As an example, we instantiate a single Graphical Lasso problem (see the
problem formulation below). We input the empirical covariance matrix
\texttt{S} and the number of samples \texttt{N}. We can choose to model
latent variables and set the regularization parameters via the other
input arguments.

\begin{Shaded}
\begin{Highlighting}[]
\CommentTok{# Import the main class of the package}
\ImportTok{from}\NormalTok{ gglasso.problem }\ImportTok{import}\NormalTok{ glasso_problem}

\CommentTok{# Define a SGL problem instance with given data S}
\NormalTok{problem  }\OperatorTok{=}\NormalTok{ glasso_problem(S, N, reg }\OperatorTok{=} \VariableTok{None}\NormalTok{,}
\NormalTok{                          reg_params }\OperatorTok{=}\NormalTok{ \{}\StringTok{'lambda1'}\NormalTok{: }\FloatTok{0.01}\NormalTok{\}, latent }\OperatorTok{=} \VariableTok{False}\NormalTok{)}
\end{Highlighting}
\end{Shaded}

As a second example, we instantiate a Group Graphical Lasso problem with
latent variables. Typically, the optimal choice of the regularization
parameters are not known and are determined via model selection.

\begin{Shaded}
\begin{Highlighting}[]
\CommentTok{# Define a GGL problem instance with given data S}
\NormalTok{problem  }\OperatorTok{=}\NormalTok{ glasso_problem(S, N, reg }\OperatorTok{=} \StringTok{"GGL"}\NormalTok{, reg_params }\OperatorTok{=} \VariableTok{None}\NormalTok{, latent }\OperatorTok{=} \VariableTok{True}\NormalTok{)}
\end{Highlighting}
\end{Shaded}

Depending on the input arguments, \texttt{glasso\_problem} comprises two
main modes:

\begin{itemize}
\tightlist
\item
  if regularization parameters are specified, the problem-dependent
  default solver is called.
\item
  if regularization parameters are \emph{not} specified,
  \texttt{GGLasso} performs model selection via grid search and the
  extended BIC criterion (Foygel and Drton 2010)).
\end{itemize}

\begin{Shaded}
\begin{Highlighting}[]
\NormalTok{problem.solve()}
\NormalTok{problem.model_selection()}
\end{Highlighting}
\end{Shaded}

For further information on the input arguments and methods, we refer to
the
\href{https://gglasso.readthedocs.io/en/latest/problem-object.html}{detailled
documentation}.

\begin{figure}
\centering
\includegraphics[width=0.9\textwidth,height=\textheight]{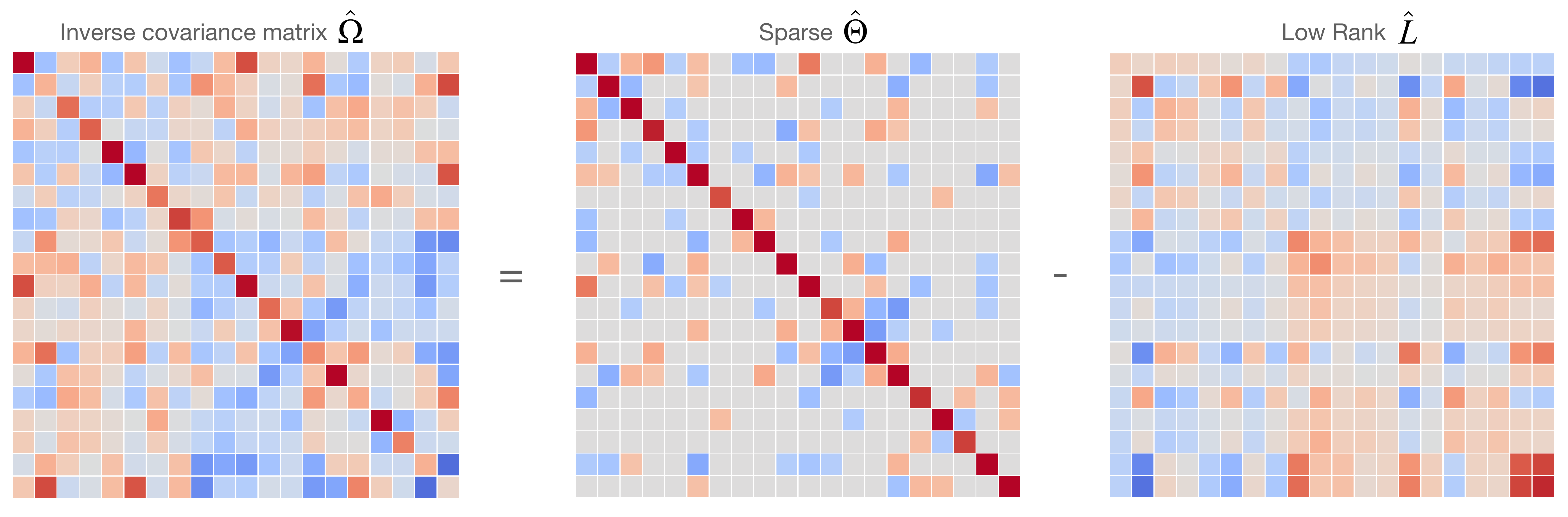}
\caption{Illustration of the latent SGL: The estimated inverse
covariance matrix \(\hat \Omega\) decomposes into a sparse component
\(\hat \Theta\) (central) and a low-rank component \(\hat L\) (right).
\label{fig1}}
\end{figure}

\hypertarget{problem-formulation}{%
\subsection{Problem formulation}\label{problem-formulation}}

We list important special cases of the problem formulation given in
\autoref{eq:problem}. For a mathematical formulation of each special
case, we refer to the
\href{https://gglasso.readthedocs.io/en/latest/math-description.html}{documentation}.

\hypertarget{SGL}{%
\subsubsection{\texorpdfstring{Single Graphical Lasso
(\emph{SGL}):}{Single Graphical Lasso (SGL):}}\label{SGL}}

For \(K=1\), the problem reduces to the single (latent variable)
Graphical Lasso where \[
\mathcal{P}(\Theta) = \lambda_1 \sum_{i \neq j} |\Theta_{ij}|.
\] An illustration of the single latent variable Graphical Lasso model
output is shown in \autoref{fig1}.

\hypertarget{GGL}{%
\subsubsection{\texorpdfstring{Group Graphical Lasso
(\emph{GGL}):}{Group Graphical Lasso (GGL):}}\label{GGL}}

For \[
\mathcal{P}(\Theta) = \lambda_1 \sum_{k=1}^{K} \sum_{i \neq j} |\Theta_{ij}^{(k)}| + \lambda_2  \sum_{i \neq j} \left(\sum_{k=1}^{K} |\Theta_{ij}^{(k)}|^2 \right)^{\frac{1}{2}}
\] we obtain the Group Graphical Lasso as formulated in (Danaher, Wang,
and Witten 2013).

\hypertarget{FGL}{%
\subsubsection{\texorpdfstring{Fused Graphical Lasso
(\emph{FGL}):}{Fused Graphical Lasso (FGL):}}\label{FGL}}

For \[
\mathcal{P}(\Theta) = \lambda_1 \sum_{k=1}^{K} \sum_{i \neq j} |\Theta_{ij}^{(k)}| + \lambda_2  \sum_{k=2}^{K}   \sum_{i \neq j} |\Theta_{ij}^{(k)} - \Theta_{ij}^{(k-1)}|
\] we obtain Fused (also called Time-Varying) Graphical Lasso (Danaher,
Wang, and Witten 2013; Tomasi et al. 2018; Hallac et al. 2017).

\hypertarget{nonconforming-ggl}{%
\subsubsection{Nonconforming GGL:}\label{nonconforming-ggl}}

Consider the GGL case in a situation where not all variables are
observed in every instance \(k=1,\dots,K\). \texttt{GGLasso} is able to
solve these problems and include latent variables. We provide the
mathematical details in the
\href{https://gglasso.readthedocs.io/en/latest/math-description.html\#ggl-the-nonconforming-case}{documentation}
and give an
\href{https://gglasso.readthedocs.io/en/latest/auto_examples/plot_nonconforming_ggl.html\#sphx-glr-auto-examples-plot-nonconforming-ggl-py}{example}.

\hypertarget{optimization-algorithms}{%
\subsection{Optimization algorithms}\label{optimization-algorithms}}

The \texttt{GGLasso} package implements several methods with provable
convergence guarantees for solving the optimization problems formulated
above.

\begin{itemize}
\item
  \emph{ADMM}: for all problem formulations we implemented the ADMM
  algorithm (Boyd et al. 2011). ADMM is a flexible and efficient
  optimization scheme which is specifically suited for Graphical Lasso
  problems as it only relies on efficient computation of the proximal
  operators of the involved functions (Danaher, Wang, and Witten 2013;
  Tomasi et al. 2018; Ma, Xue, and Zou 2013).
\item
  \emph{PPDNA}: for GGL and FGL problems without latent variables, we
  included the proximal point solver proposed in (Zhang et al. 2019,
  2020). According to the numerical experiments in (Zhang et al. 2020),
  PPDNA can be an efficient alternative to ADMM especially for fast
  local convergence.
\item
  \emph{block-ADMM}: for SGL problems without latent variables, we
  implemented a method which solves the problem block-wise, following
  the proposal in (Witten, Friedman, and Simon 2011). This wrapper
  simply applies the ADMM solver to all connected components of the
  empirical covariance matrix after thresholding.
\end{itemize}

\begin{figure}
\centering
\includegraphics[width=0.9\textwidth,height=\textheight]{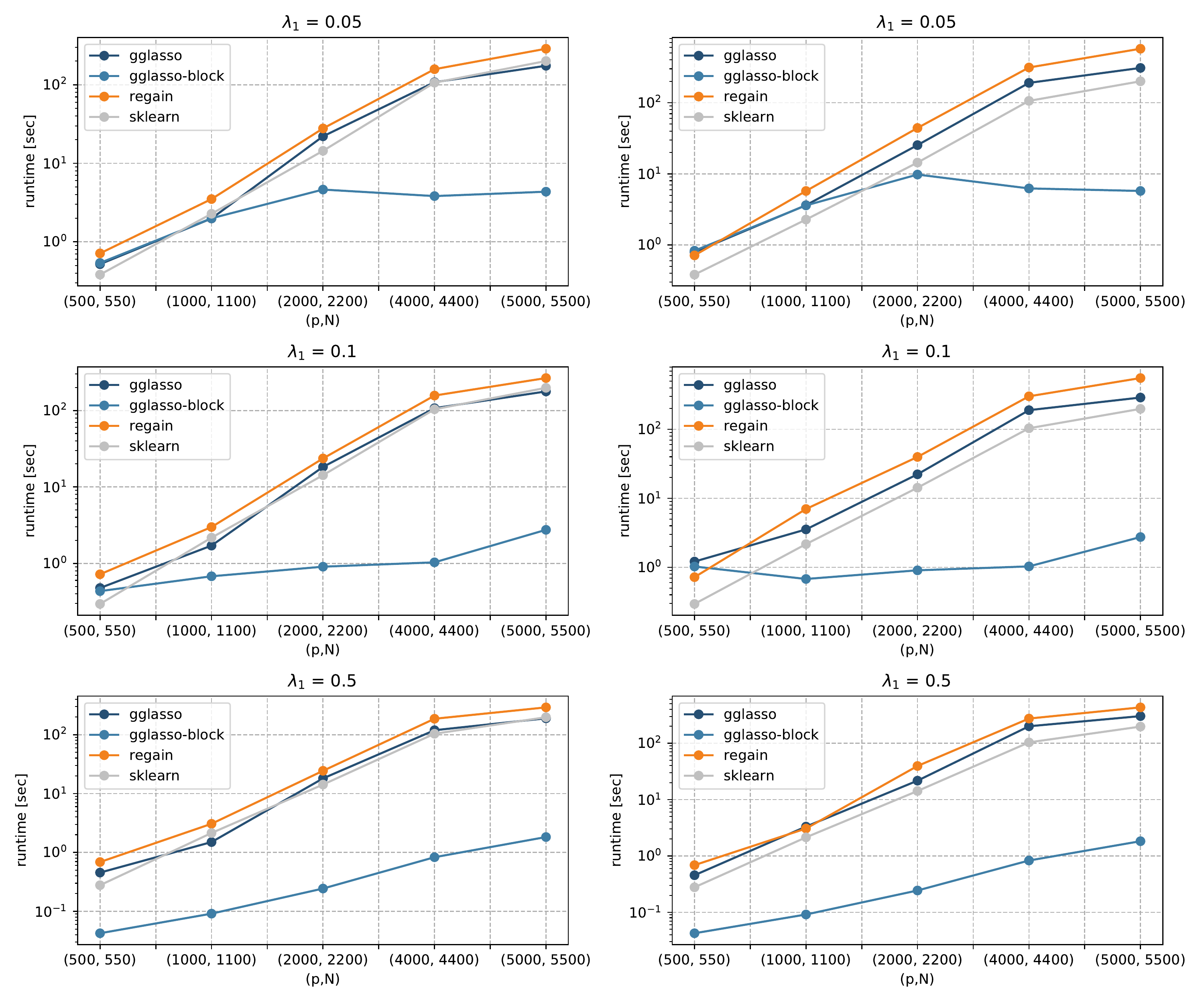}
\caption{Runtime comparison for SGL problems of varying dimension and
sample size at three different \(\lambda_1\) values. The left column
shows the runtime at low accuracy, the right column at high accuracy.
\label{fig2}}
\end{figure}

\hypertarget{benchmarks-and-applications}{%
\subsection{Benchmarks and
applications}\label{benchmarks-and-applications}}

In our example gallery, we included benchmarks comparing the solvers in
\texttt{GGLasso} to state-of-the-art software as well as illustrative
examples explaining the usage and functionalities of the package. We
want to emphasize the following examples:

\begin{itemize}
\item
  \href{https://gglasso.readthedocs.io/en/latest/auto_examples/plot_benchmarks.html\#sphx-glr-auto-examples-plot-benchmarks-py}{Benchmarks}
  for SGL problems: our solver is competitive with \texttt{scikit-learn}
  and \texttt{regain}. The newly implemented block-wise solver is highly
  efficient for large sparse networks (see \autoref{fig2} for runtime
  comparison at
  \href{https://gglasso.readthedocs.io/en/latest/auto_examples/plot_benchmarks.html\#calculating-the-accuracy}{low
  and high accuracy}, respectively).
\item
  \href{https://gglasso.readthedocs.io/en/latest/auto_examples/plot_soil_example.html\#sphx-glr-auto-examples-plot-soil-example-py}{Soil
  microbiome application}: following (Kurtz, Bonneau, and Müller 2019),
  we demonstrate how latent variables can be used to identify hidden
  confounders in microbial network inference.
\item
  \href{https://gglasso.readthedocs.io/en/latest/auto_examples/plot_nonconforming_ggl.html\#sphx-glr-auto-examples-plot-nonconforming-ggl-py}{Nonconforming
  GGL}: we illustrate how to use \texttt{GGLasso} for GGL problems with
  missing variables.
\end{itemize}

\hypertarget{acknowledgements}{%
\section{Acknowledgements}\label{acknowledgements}}

We thank Prof.~Dr.~Michael Ulbrich, TU Munich, for supervising the
Master's thesis of FS that led to the development of the software. We
also thank Dr.~Zachary D. Kurtz for helping with testing of the latent
graphical model implementation.

\hypertarget{references}{%
\section*{References}\label{references}}
\addcontentsline{toc}{section}{References}

\hypertarget{refs}{}
\leavevmode\hypertarget{ref-Banerjee2008}{}%
Banerjee, Onureena, Laurent El Ghaoui, and Alexandre D'Aspremont. 2008.
``Model selection through sparse maximum likelihood estimation for
multivariate gaussian or binary data.'' \emph{J. Mach. Learn. Res.} 9:
485--516. \url{https://doi.org/10.5555/1390681.1390696}.

\leavevmode\hypertarget{ref-Boyd2011}{}%
Boyd, Stephen, Neal Parikh, Eric Chu, Borja Peleato, and Jonathan
Eckstein. 2011. ``Distributed Optimization and Statistical Learning via
the Alternating Direction Method of Multipliers.'' \emph{Found. Trends
Mach. Learn.} 3 (1). Hanover, MA, USA: Now Publishers Inc.: 1--122.
\url{https://doi.org/10.1561/2200000016}.

\leavevmode\hypertarget{ref-Chandrasekaran2012}{}%
Chandrasekaran, Venkat, Pablo A. Parrilo, and Alan S. Willsky. 2012.
``Latent Variable Graphical Model Selection via Convex Optimization.''
\emph{Ann. Statist.} 40 (4). Institute of Mathematical Statistics:
1935--67. \url{https://doi.org/10.1214/11-aos949}.

\leavevmode\hypertarget{ref-Danaher2013}{}%
Danaher, Patrick, Pei Wang, and Daniela M. Witten. 2013. ``The Joint
Graphical Lasso for Inverse Covariance Estimation Across Multiple
Classes.'' \emph{J. R. Stat. Soc. B} 76 (2). Wiley: 373--97.
\url{https://doi.org/10.1111/rssb.12033}.

\leavevmode\hypertarget{ref-Dempster:1972}{}%
Dempster, Arthur P. 1972. ``Covariance selection.'' \emph{Biometrics} 28
(1). JSTOR: 157--75. \url{https://doi.org/10.2307/2528966}.

\leavevmode\hypertarget{ref-Foygel2010}{}%
Foygel, Rina, and Mathias Drton. 2010. ``Extended Bayesian Information
Criteria for Gaussian Graphical Models.'' In \emph{Advances in Neural
Information Processing Systems}, edited by J. Lafferty, C. Williams, J.
Shawe-Taylor, R. Zemel, and A. Culotta. Vol. 23. Curran Associates, Inc.
\url{https://proceedings.neurips.cc/paper/2010/file/072b030ba126b2f4b2374f342be9ed44-Paper.pdf}.

\leavevmode\hypertarget{ref-Friedman2007}{}%
Friedman, J., T. Hastie, and R. Tibshirani. 2007. ``Sparse Inverse
Covariance Estimation with the Graphical Lasso.'' \emph{Biostatistics} 9
(3). Oxford University Press (OUP): 432--41.
\url{https://doi.org/10.1093/biostatistics/kxm045}.

\leavevmode\hypertarget{ref-Hallac2017}{}%
Hallac, David, Youngsuk Park, Stephen Boyd, and Jure Leskovec. 2017.
``Network Inference via the Time-Varying Graphical Lasso.'' In
\emph{Proceedings of the 23rd ACM SIGKDD International Conference on
Knowledge Discovery and Data Mining}. ACM.
\url{https://doi.org/10.1145/3097983.3098037}.

\leavevmode\hypertarget{ref-Kurtz2019}{}%
Kurtz, Zachary D, Richard Bonneau, and Christian L Müller. 2019.
``Disentangling microbial associations from hidden environmental and
technical factors via latent graphical models.'' \emph{bioRxiv},
2019.12.21.885889. \url{https://doi.org/10.1101/2019.12.21.885889}.

\leavevmode\hypertarget{ref-Laska2017}{}%
Laska, Jason, and Manjari Narayan. 2017. ``skggm 0.2.7: A scikit-learn
compatible package for Gaussian and related Graphical Models,'' July.
Zenodo. \url{https://doi.org/10.5281/zenodo.830033}.

\leavevmode\hypertarget{ref-Ma2013}{}%
Ma, Shiqian, Lingzhou Xue, and Hui Zou. 2013. ``Alternating Direction
Methods for Latent Variable Gaussian Graphical Model Selection.''
\emph{Neural Comput.} 25 (8). MIT Press - Journals: 2172--98.
\url{https://doi.org/10.1162/neco_a_00379}.

\leavevmode\hypertarget{ref-Pedregosa2011}{}%
Pedregosa, F., G. Varoquaux, A. Gramfort, V. Michel, B. Thirion, O.
Grisel, M. Blondel, et al. 2011. ``Scikit-Learn: Machine Learning in
Python.'' \emph{J Mach Learn Res} 12: 2825--30.

\leavevmode\hypertarget{ref-Tomasi2018}{}%
Tomasi, Federico, Veronica Tozzo, Saverio Salzo, and Alessandro Verri.
2018. ``Latent Variable Time-Varying Network Inference.'' In
\emph{Proceedings of the 24th ACM SIGKDD International Conference on
Knowledge Discovery \& Data Mining}. ACM.
\url{https://doi.org/10.1145/3219819.3220121}.

\leavevmode\hypertarget{ref-Witten2011}{}%
Witten, Daniela M., Jerome H. Friedman, and Noah Simon. 2011. ``New
Insights and Faster Computations for the Graphical Lasso.'' \emph{J.
Comput. Graph. Statist.} 20 (4). Taylor \& Francis: 892--900.
\url{https://doi.org/10.1198/jcgs.2011.11051a}.

\leavevmode\hypertarget{ref-Yuan2007}{}%
Yuan, M., and Y. Lin. 2007. ``Model Selection and Estimation in the
Gaussian Graphical Model.'' \emph{Biometrika} 94 (1). Oxford University
Press (OUP): 19--35. \url{https://doi.org/10.1093/biomet/asm018}.

\leavevmode\hypertarget{ref-Zhang2019}{}%
Zhang, Ning, Yangjing Zhang, Defeng Sun, and Kim-Chuan Toh. 2019. ``An
Efficient Linearly Convergent Regularized Proximal Point Algorithm for
Fused Multiple Graphical Lasso Problems,'' February.
\url{http://arxiv.org/abs/1902.06952v1}.

\leavevmode\hypertarget{ref-Zhang2020}{}%
Zhang, Yangjing, Ning Zhang, Defeng Sun, and Kim-Chuan Toh. 2020. ``A
Proximal Point Dual Newton Algorithm for Solving Group Graphical Lasso
Problems.'' \emph{SIAM J. Optim.} 30 (3): 2197--2220.
\url{https://doi.org/10.1137/19M1267830}.

\end{document}